# Contrasting dynamics of seed banks and standing vegetation of annuals and perennials along a rainfall gradient


Niv DeMalach[1,2*], Jaime Kigel[2], Marcelo Sternberg[1]

[1] School of Plant Sciences and Food Security, George S. Wise Faculty of Life Sciences, Tel Aviv University, Tel Aviv, Israel

[2] Institute of Plant Sciences and Genetics in Agriculture, Robert H. Smith Faculty of Agriculture, Food and Environment, Hebrew University of Jerusalem, Rehovot, Israel

* Corresponding author:

Niv.demalach@mail.huji.ac.il

https://orcid.org/0000-0002-4509-5387

https://**orcid**.org/0000-0001-8710-4141 (Marcelo Sternberg)

https://orcid.org/0000-0003-3028-9522 (Jaime Kigel)








**ABSTRACT**

The soil seed bank is a major component of plant communities. However, long-term analyses of the dynamics of the seed bank and the ensuing vegetation are rare. Here, we studied the dynamics in plant communities with high dominance of annuals in Mediterranean, semiarid, and arid ecosystems for nine consecutive years. For annuals, we hypothesized that the density of the seed bank would be more stable than the density of the standing herbaceous vegetation. Moreover, we predicted that differences in temporal variability between the seed bank and the vegetation would increase with aridity, where year-to-year rainfall variability is higher.

We found that the temporal variability at the population level (assessed as the standard deviation of the $\log_e$-transformed density) of the nine dominant annuals in each site did not differ between the seed bank and the ensuing vegetation in any of the sites.

For the total density of annuals, patterns depended on aridity. In the Mediterranean site, the temporal variability was similar in the seed bank and the vegetation (0.40 vs. 0.40). Still, in the semiarid and arid sites, variability in the seed bank was lower than in the vegetation (0.49 vs. 1.01 and 0.63 vs. 1.38, respectively). This difference between the population-level patterns and the total density of annuals can be related to the lower population synchrony in their seed bank. In contrast, for the herbaceous perennials (all species combined), the seed bank variability was higher than in the vegetation. Overall, our results highlight the role of the seed bank in buffering the annual vegetation density with increasing climatic uncertainty typical in aridity gradients. This role is crucial under the increasing uncertainty imposed by climatic change in the region.





## 1. INTRODUCTION

Fluctuations in environmental conditions often lead to temporal fluctuations in plant populations and communities (Blüthgen et al., 2016; Hallett et al., 2018; Venable and Kimball, 2012). The soil seed bank can decrease the magnitude of density fluctuations because seed dormancy and longevity allow species to 'escape' years of stressful conditions (Gremer et al., 2016; Kigel, 1995; Thompson, 1987). The seed bank is especially important for annual plants that establish from seeds each growing season (Cohen, 1966; Eskelinen et al., 2021; Loydi and Collins, 2021; MacDonald and Watkinson, 1981; Pake and Venable, 1996). Furthermore, ecological theory predicts that the role of the seed bank in buffering population dynamics should increase with increasing uncertainty in environmental conditions (Bernhardt et al., 2020; Chesson et al., 2004; Cohen, 1966; MacDonald and Watkinson, 1981; Venable and Brown, 1988; Wisnoski and Shoemaker, 2021)

Many studies have investigated the drivers of temporal variability of community density and biomass in the vegetation (Avolio et al., 2020; Bai et al., 2004; Blüthgen et al., 2016; Hautier et al., 2015, 2014; Isbell et al., 2009; Lepš et al., 2018; Valencia et al., 2020). In contrast, the temporal variability of seed banks has been much less investigated (DeMalach et al., 2021; Levassor et al., 1990; Venable and Kimball, 2012). Here, we studied the long-term density dynamics in both the seed bank and the standing vegetation of annual and perennial herbaceous plants along a rainfall gradient in Israel for nine consecutive years. This rainfall gradient can be viewed as a gradient of increasing uncertainty because variability in rainfall increases with aridity (Cohen 1971, Sternberg et al. 2011, Tielbörger et al. 2014).

We investigated the temporal variability in density at the population level (focusing on dominant species) and community levels (density of all species combined). Importantly, fluctuations in the community level are not a simple average of the fluctuation of the





populations. First, population synchrony, the degree of correlation among species-level fluctuations, increases the community-level variability and vice versa for asynchrony (de Mazancourt et al., 2013; Yachi and Loreau, 1999). Additionally, more diverse communities are less variable due to the portfolio effect since the sum of randomly varying items is less variable than the average item ( de Mazancourt et al., 2013; Doak et al., 1998; Thibaut and Connolly, 2013; Tilman, 1999).

Our analysis aimed to test the following predictions: (1) The ratio between the densities of seeds and plants in the seed bank and in the ensuing vegetation (hereafter, density ratio), respectively, will increase with aridity, where conditions for growth and survival are more stressful. (2) For annual plants, the temporal variability in the density (of populations and the entire community) in the seed bank would be lower than in the vegetation. (3) The differences between the seed bank and the vegetation in temporal variability would increase with aridity due to the expected higher rainfall variability. (4) For perennials with a longer generation time, fluctuation in the vegetation would be smaller than seed bank fluctuation.

## 2. METHODS

### 2.1 Study sites

The study was conducted from 2001 to 2010 at three sites along a rainfall gradient in Israel (also described in Harel et al., 2011)). All sites were located over the same calcareous bedrock on south-facing slopes and experienced a similar range of mean annual temperatures (17.7 – 19.1 °C). Additionally, all sites are characterized by rainy and moderately cold winters and long dry, hot summers. Still, the three sites represent three different ecosystem types: Mediterranean (Matta LTER; N 31º 42'; E 35º 03'), semiarid (Lahav, N 31º 23'; E 34º 54'), and arid (Sde-Boker, N 30º 52', E 34º 46'). The long-term mean annual rainfall in these





three sites is 540, 300, and 90 mm, with a coefficient of variation (CV) of 30%, 37%, and 51%, respectively.

All sites are characterized by a mosaic of open herbaceous patches and semi-deciduous shrubs (Mostly *Sarcopoterium spinosum* in the Mediterranean and semiarid sites and *Zygophyllum dumosum* in the arid site). However, our analysis focuses only on open patches (we avoided areas with shrubs). In all sites, annuals dominate over perennial herbs in the herbaceous patches in terms of biomass, density, and richness (Tielbörger et al. 2014). Germination of annuals and re-growth of most perennials occur soon after the first major rainfall events (~10–20 mm). The length of the typical rainy season increases from the desert (December to March) to the Mediterranean (October to May). Previous studies at the sites have shown that for annuals, the risk of reproductive failure (germination without reaching reproduction) is high at the arid site, low at the semiarid, and negligible at the Mediterranean site (Petrů and Tielbörger, 2008). Each site included five plots of 250 m$^2$ (10 m × 25 m) with a distance of ~20m between adjacent plots.

The three sites were fenced against grazing (by sheep and goats) in 2001. Before establishing the experimental plots, grazing intensity was high in the semiarid site, intermediate in the Mediterranean, and negligible in the arid site (M. Sternberg, personal observations). Accordingly, fencing led to a strong compositional shift in the semiarid site, a weaker trend in the Mediterranean site, and no directional trend in the arid site (DeMalach et al., 2021).

## 2.2 Vegetation and seed bank sampling

Soil seed bank samples were collected each year (2001–2009) in September before the onset of the rainy season. The late sampling date ensured that seeds present in these soil samples were exposed for at least five months to local natural climatic conditions after seed set and shedding (March/April). This period of field exposure is essential for breaking seed dormancy in some species, particularly in regions with hot and dry summers, leading to





conditions of reducing dormancy and enhancing germination at the onset of the rainy season.

Moreover, since significant losses due to granivory (mainly by ants and rodents) occur during

this period, collecting at this late date ensures that the number of seeds in the soil samples

represents a more accurate representation of potential seed germination at each station

(Lebrija-Trejos et al. 2011). Ten randomly-located samples (soil cores) were taken in the

open herbaceous dominant patches (i.e., patches without shrub cover) in each of the five plots

over an area of 5×5 cm and a depth of 5 cm, including the surface litter. Each sample was

brought to the lab and thoroughly mixed while removing stones and coarse roots. The soil

and plant litter containing seeds were spread in plastic trays that enabled drainage (12×14 cm,

6.5 cm depth). The trays were irrigated during winter (beginning early October) in a net

house at the Botanical Garden of Tel Aviv University. Emerging seedlings were identified,

counted, and continuously removed until no further emergence was observed (mid-March).

Each soil sample's overall germinable seed bank was assessed by repeating the germination

procedure for each tray for three consecutive growing seasons (winters) to account for seed

dormancy. The seed density of each soil sample was estimated as the sum of germinated

seedlings over the three emergence trial seasons. Seed bank trays were naturally dried in the

net house during summer to mimic typical hot, dry field conditions. At the end of the third

season, each soil sample was passed through 5- and 0.30-mm sieves to retrieve non-

germinated seeds that were counted under a microscope (80× magnification). Since the

number of retrieved non-germinated seeds was very low (<1% of the total number of

emerged seedlings) and the procedure was very time-consuming, this fraction of the seed

bank was not considered in further analyses (see Harel et al. 2011).

The annual sampling of the vegetation was conducted at peak biomass (March–April), during

the growing seasons of 2000/2001 to 2009/2010 (except in 2004/2005). Ten random samples

(20 x 20 cm quadrats) were taken in each of the five plots. Plant roots were cut just below the





soil surface to prevent plant fragmentation, thus allowing the counting of individual plants of each species in the lab.

Hereafter, we refer to the growing season in relation to the timing of the vegetation sampling rather than the seed bank sampling (e.g., the growing season of 2009/2010 will be referred to as '2010' for both the seed bank and the vegetation, despite the seed bank being sampled during September 2009).

## 2.3 Statistical analyses

Studies of temporal variability of aboveground vegetation often use permanent plots where the same area is sampled yearly (Valencia et al., 2020). Permanent plots enable disentangling temporal variability from the sampling error caused by spatial heterogeneity. Here, we could not use permanent plots because we are unaware of any non-destructive approach for sampling the soil seed bank. Therefore, we also applied a destructive sampling for the vegetation. Still, we aimed to quantify the actual year-to-year fluctuations while minimizing the effects of all sources of sampling error, which can artificially inflate temporal fluctuations (Kalyuzhny et al. 2014). Hence, to reduce this bias, we aggregated (averaged) all samples from the same year (10 quadrates × 5 plots). We have found that 50 samples are sufficient to account for the considerable noise caused by the high spatial heterogeneity in our research sites (DeMalach et al., 2021).

The density of each species in the seed bank and in the vegetation, i.e., the number of individual plants and seeds per square meter, was calculated by multiplying the average number of individuals per sampling unit in the vegetation and the seed bank by 25 and 400, respectively (the sampling unit's area was 1/25 m$^2$ for vegetation and 1/400 m$^2$ for soil).





At the population level, in each site, we analyzed the nine most dominant species in terms of relative abundance (all annuals) to avoid sampling errors due to rare species. For these analyses, we added one to all observations to avoid zeros (which are problematic when data is shown on a log scale or log-transformed). In the second stage, data of the total density (all species combined) was calculated separately for the annual species and the herbaceous perennials (geophytes and hemicryptophytes).

We quantified temporal variability in density using the coefficient of variation (CV) and the standard deviation of the $\log_e$-transformed data (hereafter $SD_{log}$). As noted by Lewontin (1966), both measures are invariant to units or any multiplicate change (when all values are multiplied by a constant), but $SD_{log}$ enables more powerful statistical tests. Differences in CV between the vegetation and the seed bank were tested using the standard asymptotic test (Feltz and Miller, 1996), which assumes a normal distribution. For $SD_{log}$ we applied Levene's test (Levene, 1960) on the log-transformed data (i.e., assuming log-normal distribution, which better fits the data). As expected, the two measures showed very similar trends, but in many cases, statistical tests were significant only for $SD_{log}$.

Synchrony among the nine dominant species was calculated with the indices proposed by Loreau and de Mazancourt (2008) and by Gross et al. (2014) (hereafter 'Loreau' and 'Gross'), using the R package 'codyn' (Hallett et al., 2016). Loreau's index is standardized between 0 (perfect asynchrony) and 1 (perfect synchrony), while Gross's index is standardized between -1 (perfect asynchrony) and 1 (perfect synchrony). These indices are complementary because Loreau's index is based on species variance and is, therefore, highly influenced by relative abundance. In contrast, Gross' index is based on correlation and consequently weighs all species equally (Hallett et al., 2016). For diversity, we used the inverse Simpson's diversity index because it is more robust than species richness to the four-





times difference between the seed bank and the vegetation in the spatial scale of the sampling (Chase and Knight, 2013).

### 3. RESULTS

At the population level, there were long-term trends at the Mediterranean and semiarid sites. The dominant grasses (*Avena sterilis* and *Stipa capensis*) increased in abundance, while the other dominant species decreased or showed no long-term trends (Fig.1, 2). Nonetheless, none of the species at the arid site have shown a directional trend (Fig. 3). Surprisingly, we did not find differences in the temporal variability for any of the dominant species in all three sites (Fig. 1-3).

At the community level, we found that the total density of annuals (all species combined) was about one order of magnitude higher in the seed bank than in the vegetation (Mediterranean site: $15 \cdot 10^3$ vs. $2 \cdot 10^3$ individuals $m^{-2}$, $P < 0.001$; semiarid site: $8 \cdot 10^3$ vs. $1.5 \cdot 10^3$ individuals $m^{-2}$, $P = 0.001$; arid site: 650 vs. 40 individuals $m^{-2}$, $P < 0.001$).

In the semiarid site, the increase in density of the dominant grass was lower than the decline of the other species. Hence, there was a decline in the total density of the vegetation but not the seed bank (Fig. 4). In the Mediterranean and semiarid sites, we did not detect significant long-term trends. The vegetation's CV and $SD_{log}$ density of all annuals increased with aridity (Fig. 4). In the Mediterranean site, $SD_{log}$ was similar in the seed bank and the vegetation. However, in the semiarid and arid sites, the $SD_{log}$ of the seed bank was lower. Although patterns of $SD_{log}$ and CV were very similar (Fig. 4), differences between the vegetation and seed bank were significant only for $SD_{log}$.

The differences between the temporal variability patterns at the population level (Fig 1-3) and total density (Fig. 4) can be related to differences in synchrony. Synchrony in the vegetation was higher than in the seed bank based on Loreau's (0.78 vs. 0.31) and Gross' (0.71 vs. 0.25)





indices. In contrast, we found no differences between the seed bank and the vegetation in terms of Simpson's diversity (Appendix S1, Fig. S1), implying that the portfolio effect is probably not the driver of the higher community-level stability of the seed bank.

Perennial plants had a lower density than annuals in all sites (in the arid site, their density was too low for analyses, see methods). As for annuals, the density of perennials in the seed bank was higher than the vegetation (Mediterranean site: 550 vs. 40 individuals $m^{-2}$, P < 0.001; semiarid site: 230 vs. 50 individuals $m^{-2}$, P = 0.003). For temporal variability, we found no differences between the seed bank and the vegetation in the Mediterranean site (Fig. 5). Nonetheless, at the semiarid site, the seed bank of the perennial species had higher temporal variability than the vegetation.





## 4. DISCUSSION

The ecological theory predicts that the seed bank stabilizes annual communities, especially in an unstable environment (Cohen 1966, MacDonald and Watkinson 1981, Venable and Brown 1988, Chesson et al. 2004). Nonetheless, we did not observe higher population stability in the seed bank (compared with the vegetation) for any of the dominant species. Instead, the higher stability of the seed bank was apparent in terms of the total density of annuals. Additionally, we supported the prediction that the difference in temporal variability between the vegetation and the seed bank should increase with aridity (stronger buffering by the seed bank as aridity increases). Lastly, we showed that, unlike annuals, for perennial plants, the vegetation is more stable than the seed bank.

### 4.1 Density ratio between the seed bank and the vegetation

The density ratios between the annual seed bank and the annual vegetation varied among sites, with values of ~7, 5, and 17 in the Mediterranean, semiarid, and arid sites, respectively. This variation in density ratios could be related to differences in germination fraction. Previous analyses have demonstrated that under constant irrigation within net-house conditions, the germination fraction of species from the more arid sites is higher than the germination fraction in the other sites (Harel et al., 2011). However, the germination fraction in the arid site is expected to be lower under natural conditions because of the lower water availability, especially in dry years.

A potential driver for the higher density ratio in the arid site is variability in seedling survival. A previous study has shown that the survival probabilities in the Mediterranean and semiarid sites are similar (Metz et al., 2010) but that survival in the arid site is much lower (James et al., 2011; Shackelford et al., 2021). Lastly, seed loss between the seed set and the germination season due to granivory and seed decay is expected to decrease the ratio between the densities of the seed bank and the vegetation. Therefore, the lower ratio in the semiarid





site compared with the Mediterranean site could be related to higher seed granivory by ants in that site (Lebrija-Trejos et al., 2011).

## 4.2 Potential drivers of the stability of the seed bank of annual plants

In both the semiarid and arid sites, the seed bank of the annual plants was more stable than the vegetation. This pattern could not be explained by differences in stability in the populations as we originally predicted. One possible reason is that the theory is related to the persistence of seeds in the seed bank for a considerable length of time (mid- to long-term seed bank), while here (and in most other empirical studies), we quantified a mixture of those long-term components of the seed bank and the most recent contributions to the seed bank.

Theoretical studies have shown that the temporal stability of plant communities is highly contingent on the scale in a complex way due to dispersal between "local communities" within a "metacommunity" (Wang et al., 2015; Wang and Loreau, 2016). ). In our study, the sampling was conducted at a small grain size (see methods), but the analysis was based on aggregating data across all quadrates for any given year. Indeed, at a small scale, dispersal between microhabitats may affect stability due to sink-source dynamics and mass effect (Pulliam, 1988; Shmida and Wilson, 1985). However, by aggregating data across quadrates, we focused our analysis on the entire site (in terms of extent), so the spatial processes cannot be quantified directly.

At the community level, stability is often affected by species diversity. This mechanism, known as the portfolio effect (Doak et al., 1998), is a statistical phenomenon where the sum of randomly varying items is less variable than the average item. It is termed the portfolio effect because of the analogy between the statistical averaging of fluctuations of different species and the economic principle that more diversified investment portfolios are more stable (Tilman 1999). Unfortunately, we cannot compare richness patterns because the area of





the sampling unit varies between the vegetation and seed bank (See Methods). Yet, in terms of Simpson's diversity (driven mainly by evenness and therefore less sensitive to scale), there were no differences between the seed bank and the vegetation (Appendix S1: Fig. S1).

We suspect that the lower synchrony of the seed bank was the main reason for its stability, but these results should be interpreted with caution because we have no replications. Importantly, both long-term turnover trends and year-to-year variability affect the degree of synchrony among populations (Valencia et al., 2020). In the semiarid site, higher stability was mainly driven by a lack of long-term trend that was apparent only in the vegetation. This pattern agrees with our previous finding that compositional trends are much faster in the vegetation than in the seed bank (DeMalach et al., 2021) and with the trends of the dominant species. The decline in density in the vegetation was due to the replacement of small species with large species that reach lower density than small species (with no overall trend in total biomass production (Kigel et al., 2021). In contrast, at the arid site where there is no trend, the higher stability of the seed bank is due to buffering the year-to-year variability.

## 4.3 Concluding remarks

An open question is: what are the environmental drivers of community density in the seed bank and the vegetation? Simple regressions between total rainfall amount and density of the seed bank and the vegetation have demonstrated non-significant trends, except for a positive relationship between rainfall and seed bank density in the semiarid site (Appendix S1, Fig. S2). However, the large number of potential drivers (e.g., intra-annual and inter-annual distributions of rainfall and temperature, as well as many biotic agents) is much greater than the data time series in our study (nine years for the seed bank and eight years for the vegetation) precluding analysis of drivers of fluctuations in plant and seed bank density. Overall, our results highlight the role of the seed bank in stabilizing annual communities with





increasing climatic uncertainty. This role is crucial under the increasing uncertainty imposed by climate change in the region. Moreover, soil seed banks are expected to be the hidden stocks for future plant diversity on Earth (Ma et al., 2019; Yang et al., 2021). Lastly, our comparison of the dynamics of annual and perennial densities illustrates a general feature of plant communities in drylands. While the standing vegetation is a central stabilizing element for perennials, annual plants are stabilized by the soil seed bank.

## AUTHOR CONTRIBUTION STATEMENT

MS and JK conceived the research idea within the GLOWA Jordan River project and collected the data. ND developed the seed bank and vegetation comparison, performed the statistical analysis, and wrote the first draft of the paper. All authors substantially contributed to the writing of the manuscript.

## DECLARATIONS

**Conflict of interests** We have no conflict of interests





**ACKNOWLEDGMENTS**

We are most grateful to Claus Holzapfel, Hadas Parag, Danny Harel, and Danny Wallach for soil seed bank sampling and experimental set-up efforts and Irit Konsens for vegetation sampling. Thanks are extended to Carly Goldets for English-language editing. The study was supported by the GLOWA Jordan River project and funded by the German Federal Ministry of Education and Research (BMBF), in collaboration with the Israeli Ministry of Science and Technology (MOST). ND was supported by the Tel Aviv University Postdoctoral Fellowship

**AVAILABILITY OF DATA AND MATERIALS:**

The datasets and the code are available on FigShare:

https://figshare.com/articles/dataset/The_soil_seed_bank_reduces_density_fluctuations_with _increasing_aridity/18758741

**Figure captions**

Fig. 1: The density of the nine most dominant annual plant species of the Mediterranean site (based on relative abundance) in the seed bank (red circles) and the vegetation (green circles). SD is the standard deviation of $\log_e$-transformed data. The P-values refer to comparisons of the SDs of the seed bank and the vegetation in each site (Levene's test for homogeneity of variance on the $\log_e$-transformed data). A trend line is shown when there is a statistically significant linear trend (P < 0.05). The y-axes have a logarithmic scale.

Fig. 2: The densities of the nine most dominant annual plant species of the semiarid site (based on relative abundance) in the seed bank (red circles) and the vegetation (green circles). SD is the standard deviation of $\log_e$-transformed data. The P-values refer to comparisons of the SDs of the seed bank and the vegetation in each site (Levene's test for homogeneity of variance on the $\log_e$-transformed data). A trend line is shown when there is a statistically significant linear trend (P < 0.05). The y-axes have a logarithmic scale.

Fig. 3: The densities of the nine most dominant annual plant species of the arid site (based on relative abundance) in the seed bank (red circles) and the vegetation (green circles). SD is the standard deviation of $\log_e$-transformed data. The P-values refer to comparisons of the SDs of the seed bank and the vegetation in each site (Levene's test for homogeneity of variance on the $\log_e$-transformed data). A trend line is shown when there is a statistically significant linear trend (P < 0.05). The y-axes have a logarithmic scale.

Fig. 4: The total densities of annual plants (all species) with time and their temporal variability. The left panels show the densities in the seed bank (red circles) and the vegetation (green circles). The middle and the right panels show two measures of the temporal variability, $SD_{log}$ and CV. The P-values refer to comparisons of the $SD_{log}$s and CVs of the





seed bank and the vegetation in each site. A linear trendline appears in the left panel when there is a significant temporal trend (P<0.05). The y-axes in the left panels have a logarithmic scale.

Fig. 5: The total densities of herbaceous perennials (all species) with time and their temporal variability. The lefts panels show the densities in the seed bank (red circles) and the vegetation (green circles). The middle and the right panels show two measures of the temporal variability, $SD_{log}$ and CV. The P-values refer to comparisons of the $SD_{log}$s and CVs of the seed bank and the vegetation in each site. A linear trendline appears in the right panel where there is a significant temporal trend (P<0.05). The y-axes in the left panels have a logarithmic scale.





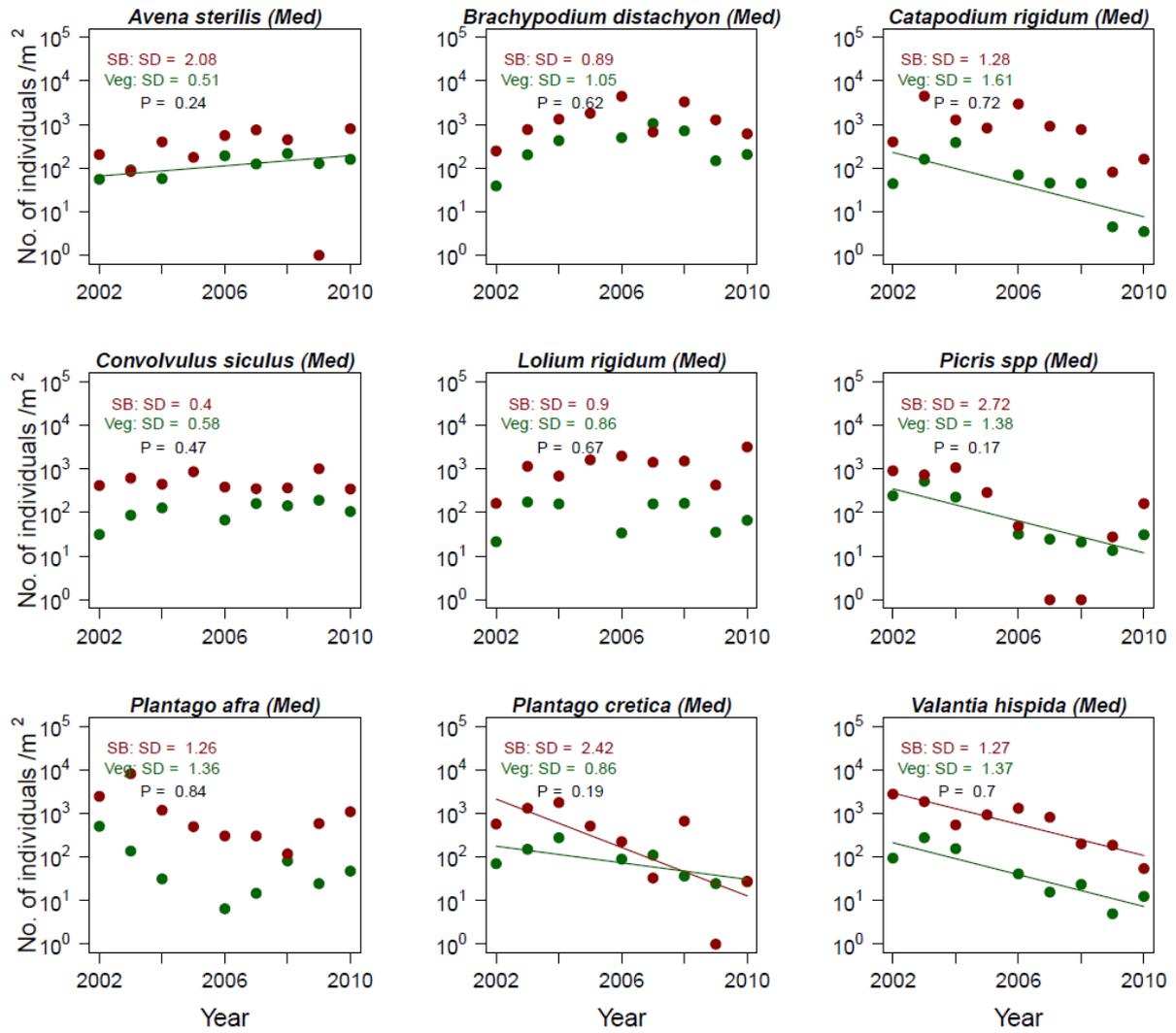

Fig. 1





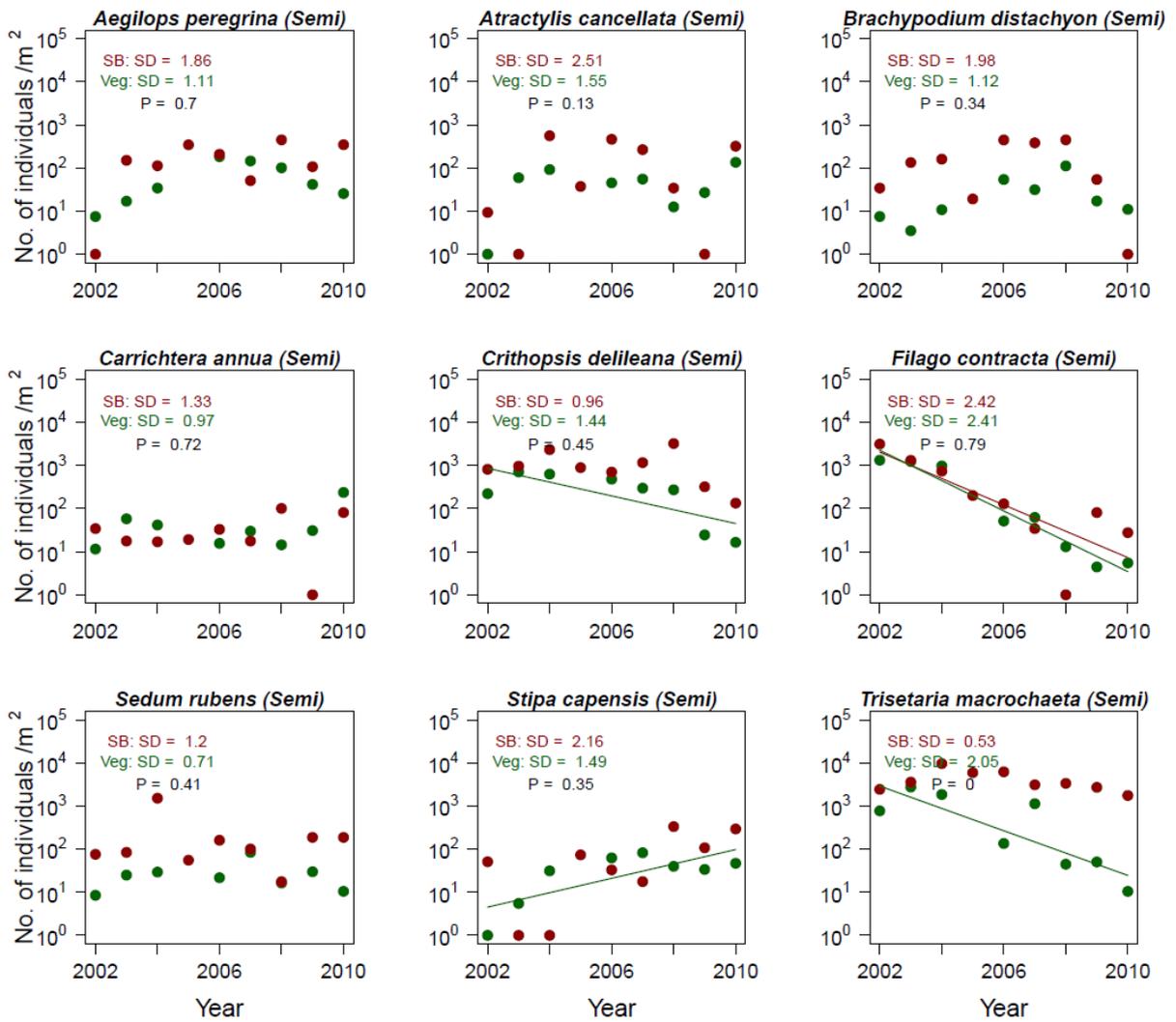

Fig. 2





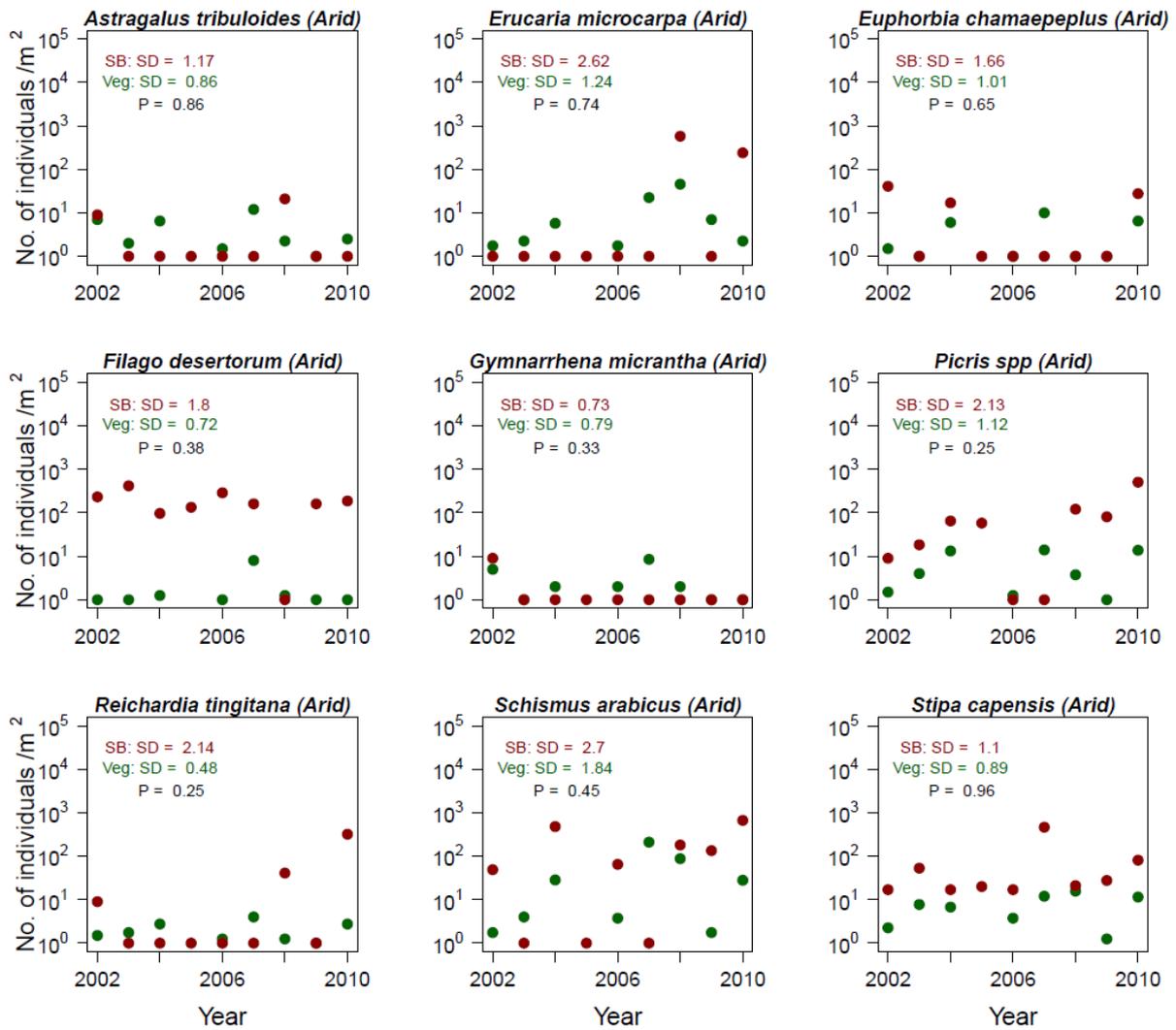

Fig. 3





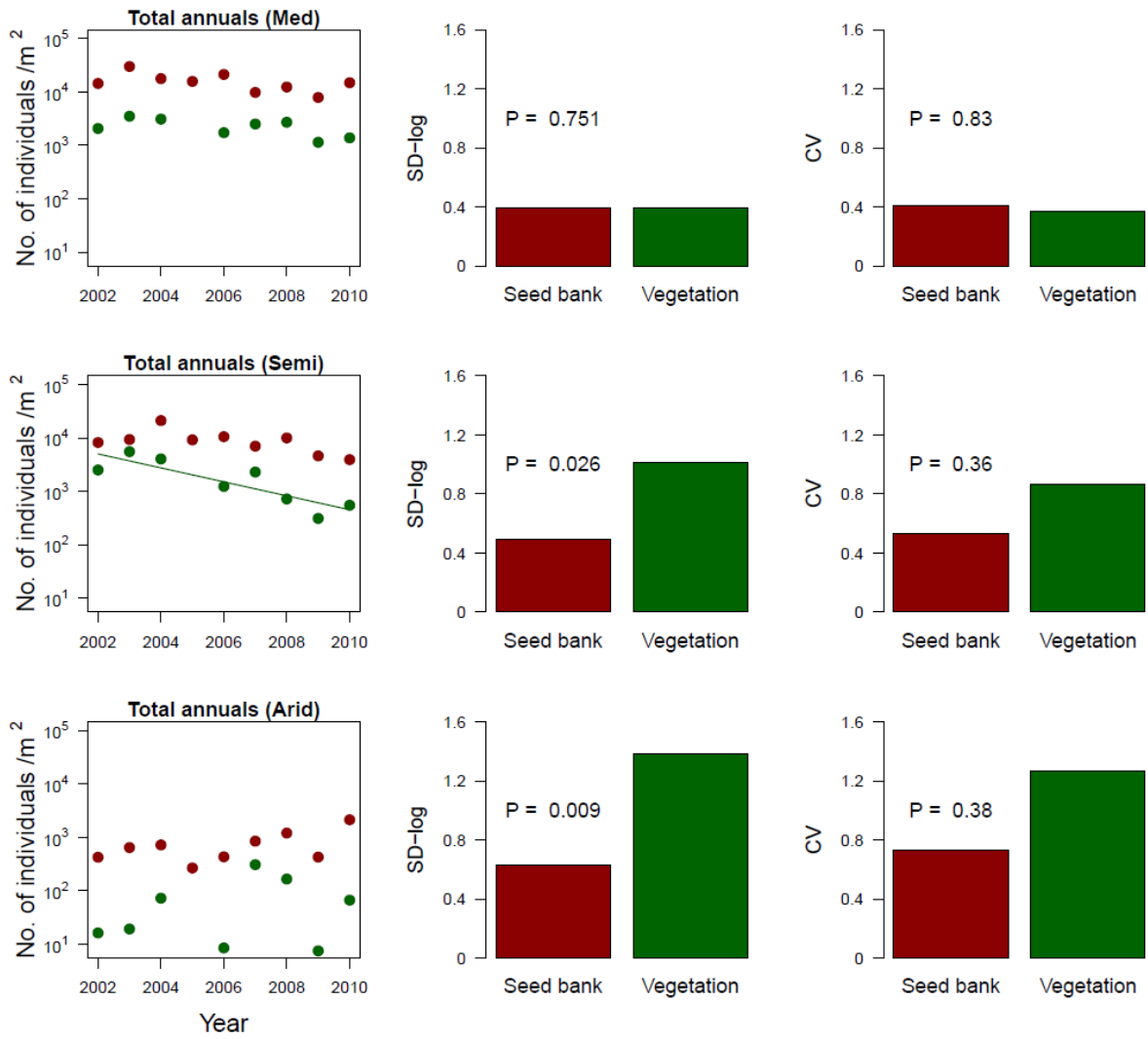

Fig. 4





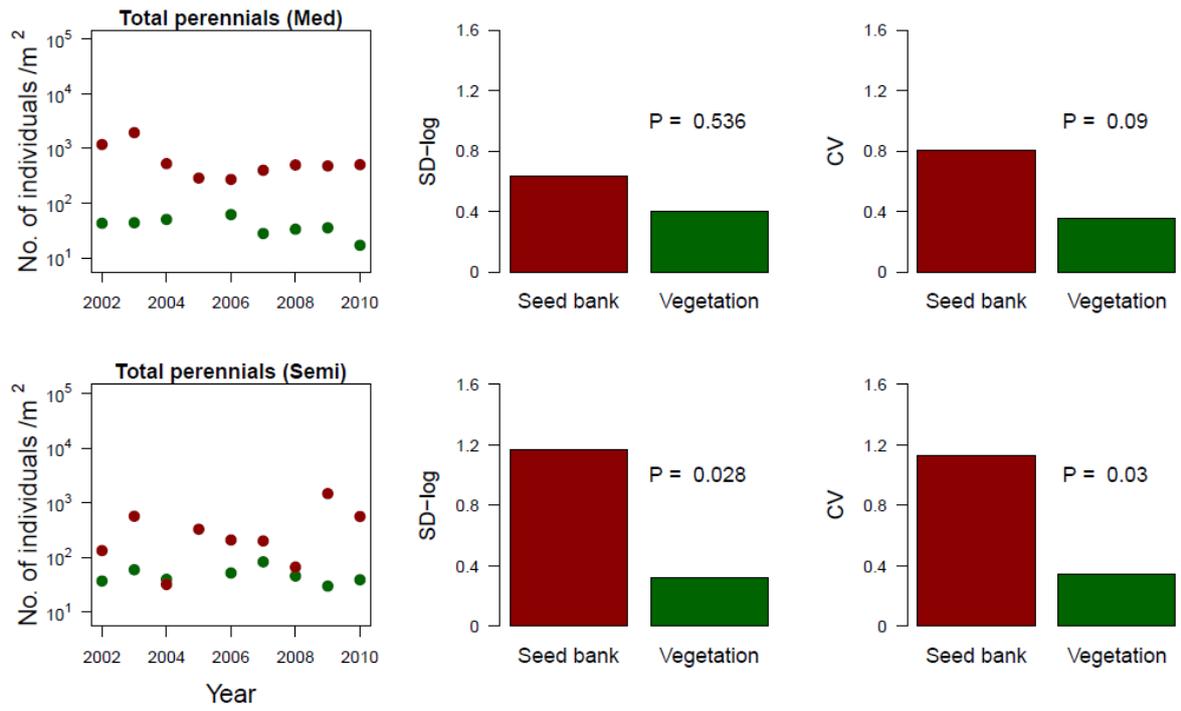

Fig. 5





Appendix S1

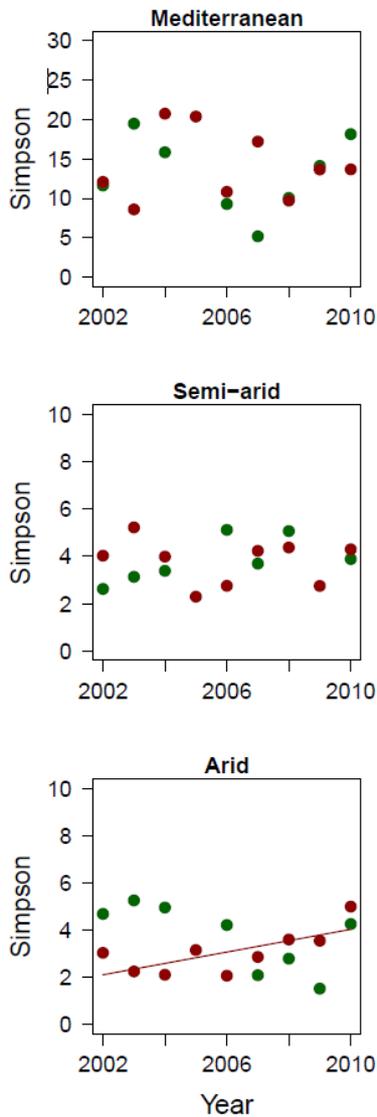

Fig. S1: Patterns of the inverse Simpson's diversity index (also known as ENS$_{PIE}$) in the seed bank (red) and the vegetation (green). Differences between the means of the seed bank and the vegetation were non-significant (P = 0.44, P = 0.21, P = 0.43 for the Mediterranean, semiarid, and arid sites). A trend line is shown when there is a statistically significant linear trend (P < 0.05). Note that the scale of the y-axes of the lower and middle panels (the semiarid and arid sites) is different than in the upper panel (the Mediterranean site).





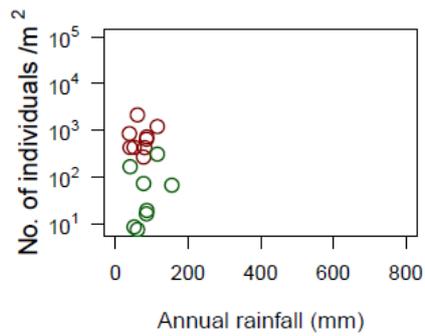

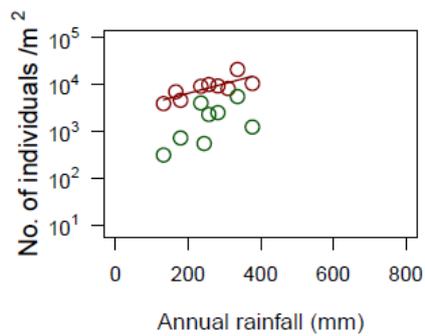

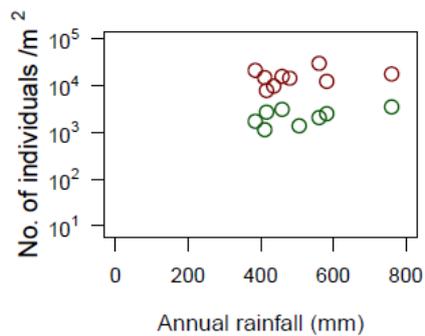

Fig. S2: The relationships between rainfall and density of annual plants in the seed bank (red circles) and in the vegetation (green). The y-axes have a logarithmic scale. For the vegetation, rainfall amount refers to the growing season before the sampling (e.g., rainfall between October 2004 and April 2005 is matched with the vegetation sampled at April 2005). Similarly, for the seed bank, rainfall amount refers to the growing season before the sampling (e.g., rainfall between October 2004 and April 2005 is matched with soil sampling in September 2006)